\documentclass[12pt]{article}
\usepackage{amsmath,amssymb,bm,graphicx}
\begin{document}

\begin{flushright}
\end{flushright}


\begin{center}
{\Large{\bf  Conditional measurement in noncontextual hidden variables models}}
\end{center}
\vskip .5 truecm
\centerline{\bf  Kazuo Fujikawa}
\vskip .4 truecm
\centerline {\it Institute of Quantum Science, College of 
Science and Technology}
\centerline {\it Nihon University, Chiyoda-ku, Tokyo 101-8308, 
Japan\footnote{Address after April 1, 2012: Mathematical Physics Laboratory, RIKEN Nishina Center, Wako, Saitama 351-0198, Japan}}

\vskip 0.5 truecm

\makeatletter
\@addtoreset{equation}{section}
\def\theequation{\thesection.\arabic{equation}}
\makeatother

\begin{abstract}
The noncontextual hidden variables
models in $d=2$, such as the ones constructed by Bell and by Kochen and Specker, have difficulties in accounting for the conditional measurement of two non-orthogonal projectors. An idea of branching in the hidden variables space, which provides a means to realize the notion of reduction effectively and describe the state preparation, is suggested as a way to resolve the difficulties associated with the conditional measurement.   
\end{abstract}


\section{Introduction}
The noncontextual hidden variables models in dimensions equal to or higher than 3 are excluded either theoretically by Gleason's theorem~\cite{gleason} or experimentally\cite{aspect} by Bell or CHSH inequalities~\cite{bell2, chsh, note}. The only $d=2$ noncontextual models such as proposed by Bell~\cite{bell1} and Kochen and Specker~\cite{kochen} have been considered to be viable models~\cite{beltrametti,peres}. Even these $d=2$ models have been shown recently to have difficulties in describing conditional measurement, if one asks a unique expression of the conditional measurement in hidden variables space~\cite{fujikawa}. If one postulates that any physical quantity should have a unique expression in hidden variables space just as any quantum mechanical quantity has a unique space-time dependence, all the noncontextual hidden variables models including $d=2$ models are thus excluded. The purpose of the present note is to show that the idea of branching 
in the hidden variables space, which is somewhat analogous to the branching in a many-worlds interpretation of quantum mechanics~\cite{everett}, saves those $d=2$ models by providing a means to describe effectively the notion of reduction and 
 state preparation and thus the conditional measurement. This modification in hidden variables models helps not to lose the entire noncontextual hidden variables models including the model of Bell in $d=2$.

In the present paper we discuss this issue on the basis of the explicit model of Bell in $d=2$~\cite{bell1} which is best known, but we also mention that the same conclusion applies to the explicit $d=2$ model by Kochen and Specker~\cite{kochen}. 

\section{Hidden variables models}

We briefly summarize the essence of hidden variables models. 
For any projector $X$ in noncontextual hidden variables models, we assign a classical number~\cite{bell1, kochen}
\begin{eqnarray}
X \rightarrow X_{\rho}(\omega)
\end{eqnarray}
where $X_{\rho}(\omega)$ assumes its eigenvalues, $X_{\rho}(\omega)=1$ or $X_{\rho}(\omega)=0$, with $\omega\in \Lambda$ standing for hidden variables. The classical number $X_{\rho}(\omega)$
depends on each given pure state $\rho$. 
For a complete set of orthogonal projection operators
$\sum_{k}X_{k}=1$,
we assume the linearity in the sense
\begin{eqnarray}
\sum_{k}X_{k \rho}(\omega)=1.
\end{eqnarray}
We then define
\begin{eqnarray}
x_{\rho}=X_{\rho}^{-1}(1)=\{\omega\in \Lambda : X_{\rho}(\omega)=1\},
\end{eqnarray}
and the probability measure associated with the projection operator $X$ is defined by 
\begin{eqnarray}
\mu[x_{\rho}]\equiv \int_{\Lambda}X_{\rho}(\omega)d\mu(\omega) =\text{Tr}[\rho X].
\end{eqnarray}
The basic assumption in hidden variables models is that one can find $X_{\rho}(\omega)$ and a measure $\mu[x_{\rho}]$ which reproduce the quantum mechanical $\text{Tr}[\rho X]$ for any $X$ and $\rho$. It is known that {\em noncontextual} hidden variables models in $d\geq3$, which are characterized by the measure $d\mu(\omega)$ independent of $X$ and $\rho$,  are excluded by Gleason's theorem~\cite{gleason, bell1, beltrametti} and the analysis of Kochen and Specker~\cite{kochen}.

The construction due to Bell in $d=2$ is based on the projector 
\begin{eqnarray}
X_{\vec{m}}=\frac{1}{2}(1+\vec{m}\cdot\vec{\sigma})
\end{eqnarray}
with $|\vec{m}|=1$, and the rule (here, $\frac{1}{2}\geq \omega\geq -\frac{1}{2}$)
\begin{eqnarray}
X_{\vec{m}\psi}(\omega)=\frac{1}{2}[1+\text{sign}(\omega+\frac{1}{2}|\vec{s}\cdot\vec{m}|)\text{sign}(\vec{s}\cdot\vec{m})]
\end{eqnarray}
for the pure state represented by the projector $|\psi\rangle\langle\psi|=\frac{1}{2}(1+\vec{s}\cdot\vec{\sigma})$ with {\em uniform noncontextual} $d\mu(\omega)=d\omega$~\cite{bell1}, namely,
\begin{eqnarray}
\int_{-\frac{1}{2}}^{\frac{1}{2}} X_{\vec{m}\psi}(\omega)d\omega=\langle \psi|X|\psi\rangle.
\end{eqnarray}
It is shown that the dispersion free $X_{\vec{m}\psi}(\omega)$ itself is not given by any density matrix parameterized by $\vec{s}$ and $\omega$~\cite{beltrametti}. We use the notation of the probability measure $\mu[x_{\psi}]$ for the present case. It has been long considered that Bell's explicit noncontextual model in $d=2$ is free from the existing no-go theorems in the framework with projectors~\cite{beltrametti, peres}.

\section{Conditional measurement}
 
In quantum mechanics one may first measure a projection operator  $B$~\cite{neumann}. Immediately after the measurement of $B$, one may measure another projector $A$. This operation is allowed even for two non-commuting projectors $[A,B]\neq 0$  
and this operation is called the conditional measurement~\cite{umegaki,davies}.
We now examine Bell's  explicit construction in $d=2$ in connection with the conditional measurement.  

One of the ways to deal with the conditional measurement on the basis of projectors is to define 
\begin{eqnarray}
\rho_{B}\equiv \frac{B\rho B}{\text{Tr}\rho B}, \hspace{1cm} \text{Tr}\rho B \neq 0,
\end{eqnarray}
then the relation 
\begin{eqnarray}
\mu[a_{\rho_{B}}]=\text{Tr}[\rho_{B}A]=\frac{\text{Tr}[(B\rho B)A]}{\text{Tr}[\rho B]}
\end{eqnarray}
holds as long as the assumed relations (2.1)-(2.4) in hidden variables models are valid for any density matrix $\rho$ which includes the density matrix $\rho_{B}$ in (3.1). This construction of (3.2) is faithful to the original quantum mechanical definition of the conditional measurement.
In Bell's construction (2.6), the projected state $\rho_{B}$ corresponds to $|\psi_{B}\rangle\langle\psi_{B}|=B$ in a matrix notation and we have the dispersion free representation (with $A=P_{\vec{m}}$, $B=P_{\vec{n}}$)
\begin{eqnarray}
A_{\psi_{B}}(\omega)=\frac{1}{2}[1+\text{sign}(\omega+\frac{1}{2}|\vec{n}\cdot\vec{m}|)\text{sign}(\vec{n}\cdot\vec{m})]
\end{eqnarray}
which is symmetric in $A$ and $B$, and we obtain the identical expression for $B_{\psi_{A}}(\omega)$.

An alternative way to analyze the conditional measurement is to {\em define} the ratio of averages~\cite{umegaki, davies}
\begin{eqnarray}
\alpha_{B}(A)=\frac{\text{Tr}\rho (BAB)}{\text{Tr}[\rho B]}, \hspace{1cm} \text{Tr}[\rho B]\neq 0
\end{eqnarray}
as the conditional probability measure of $A$ after the measurement of $B$. In (3.4), we emphasize a new composite operator $BAB$, which is no more a projection operator, while we emphasize the modification of the state in (3.2). These two are naturally identical in quantum mechanics.

For the projector in (2.5), we have
\begin{eqnarray}
P_{\vec{n}}P_{\vec{m}}P_{\vec{n}}=\frac{(1+\vec{n}\cdot\vec{m})}{2}
P_{\vec{n}}
\end{eqnarray}
and $P_{\vec{m}}P_{\vec{n}}P_{\vec{m}}=\frac{1}{2}(1+\vec{n}\cdot\vec{m})P_{\vec{m}}$. We then obtain the dispersion free representation corresponding to (3.4), 
\begin{eqnarray}
\frac{(BAB)_{\psi}(\omega)}{\langle B\rangle_{\psi}}&=&\frac{(1+\vec{n}\cdot\vec{m})}{(1+\vec{n}\cdot\vec{s})}\\
&\times&\frac{1}{2}[1+\text{sign}(\omega+\frac{1}{2}|\vec{s}\cdot\vec{n}|)\text{sign}(\vec{s}\cdot\vec{n})]\nonumber
\end{eqnarray}
using (2.6) for $B_{\psi}(\omega)$ with  $A=P_{\vec{m}}$ and $B=P_{\vec{n}}$. 

One then confirms that the conditional measurement is consistently described by either way, (3.3) or (3.6), in agreement with the prediction of quantum mechanics as
\begin{eqnarray}
\frac{\text{Tr}[\rho BAB]}{\text{Tr}[\rho B]}
=\frac{\mu[(bab)_{\psi}]}{\mu[b_{\psi}]}=\int d\omega A_{\psi_{B}}(\omega) =\frac{(1+\vec{n}\cdot\vec{m})}{2},
\end{eqnarray}
which also agrees with $\text{Tr}[\rho ABA]/\text{Tr}[\rho A]$.

This specific example in (3.7) shows that the conditional measurement in hidden variables models~\cite{bell1} does not follow the classical conditional probability rule
\begin{eqnarray}
\frac{\mu[(bab)_{\psi}]}{\mu[b_{\psi}]} \neq \frac{\mu[a_{\rho}\cap b_{\rho}]}{\mu[ b_{\rho}]}
\end{eqnarray}
for general non-commuting $A$ and $B$, despite the fact that hidden variables models are based on the dispersion free determinism. If one assumes the classical conditional probability rule on the right-hand side of (3.8) for general state $\rho$, the relation (3.7) cannot hold for non-commuting $A$ and $B$. The classical conditional probability rule (3.8), if imposed on noncontextual hidden variables models, eliminates the essential quantum mechanical properties, since 
 the right-hand side of (3.8) negates the crucial notion of reduction in quantum mechanics, as is seen by the fact that $a_{\rho}$ and $b_{\rho}$ in $\mu[a_{\rho}\cap b_{\rho}]$ are defined by the same original state $\rho$ although $\mu[a_{\rho}\cap b_{\rho}]$ is divided by $\mu[ b_{\rho}]$. 

\section{Non-uniqueness in hidden variables space} 

We recognize that the expression (3.3) and the expression (3.6)  lead to two conflicting dispersion free representations in  hidden variables space parameterized by $\omega$ for the {\em identical} quantum mechanical object\\ $\text{Tr}[\rho BAB]/\text{Tr}[\rho B]$, although both of them reproduce the same quantum mechanical result after averaging over hidden variables as in (3.7)~\cite{fujikawa}. 
We here postulate that any physical quantity should have a unique expression in hidden variables space, just as any quantum mechanical quantity has a unique space-time dependence.  This requirement is not satisfied by the expression of the 
conditional measurement in the $d=2$ noncontextual hidden variables model~\cite{bell1}.

This conflict between (3.3) and (3.6) is analogous to the case of a sum of two non-orthogonal projectors.
One may consider a linear combination of two non-collinear projectors in (2.5)
\begin{eqnarray}
E=\lambda P_{\vec{n}}+ (1-\lambda) P_{\vec{m}}, \ \ \ \  0< \lambda< 1,
\end{eqnarray}
which satisfies $0<E<1$. If one assumes that the dispersion free representation due to Bell is applied to all the operators in (4.1) separately (Bell's construction is valid for a general operator such as $E$ also), one obtains
\begin{eqnarray}
E_{\psi}(\omega)=\lambda P_{\vec{n}\psi}(\omega)+ (1-\lambda) P_{\vec{m}\psi}(\omega), \ \ \ \ 0< \lambda< 1,
\end{eqnarray}
but this is not satisfied by the positive operator  $1>E_{\psi}(\omega)>0$ on the left-hand side in the domain of the hidden variables space
with $P_{\vec{n}\psi}(\omega)=P_{\vec{m}\psi}(\omega)=0$ (or with
$P_{\vec{n}\psi}(\omega)=P_{\vec{m}\psi}(\omega)=1$). This shows that Bell's construction has an ambiguity in representing the same operator, namely, the left- and right-hand sides of (4.1), although it reproduces the result of quantum mechanics 
\begin{eqnarray}
\langle E\rangle_{\psi}=\lambda\langle  P_{\vec{n}}\rangle_{\psi}+ (1-\lambda) \langle P_{\vec{m}}\rangle_{\psi}
\end{eqnarray}
implied by (4.1) after averaging over hidden variables.

The conflict in (4.2) is essentially the original no-go theorem of von Neumann~\cite{neumann} against noncontextual hidden variables models, and its resolution is well known. One does not assign a physical significance to two incompatible operators simultaneously in hidden variables space~\cite{bell1}. One now encounters another conflict between (3.3) and (3.6) arising from non-orthogonal projectors in the analysis of the conditional measurement. 

We examine the new conflict between (3.3) and (3.6) in more detail.
From a point of view of the dual structure of operator and state $(O,\rho)$ in quantum mechanics, these two approaches are related; an extra quantum mechanical operation is included in each case, 
\begin{eqnarray}
(A,B\rho B)\ \ {\rm or}\ \ (BAB,\rho),
\end{eqnarray}
respectively, before moving to hidden variables models. These two are obviously equivalent in quantum mechanics (or in any trace representation with density matrix), but they are quite different in Bell's construction due to  the lack of definite associative properties of various operations. An interesting example is given by the measurement of $A$ immediately after the measurement of $A$. The prescription in (3.3) gives an $\omega$ independent unit representation, while the formula (3.4) with (3.6) gives $A_{\psi}(\omega)/\int A_{\psi}(\omega)d\omega$ which has the same $\omega$ dependence as the first measurement of $A$.
 
The basic issue here is how to implement the notion of reduction (or rather {\em its equivalent}) in hidden variables models. The construction $\mu[a_{\rho_{B}}]$ (3.2) and also the expression (3.3) rely directly on the notion of reduction of states caused by a measurement of $B$, which is a characteristically quantum mechanical notion foreign to deterministic realism. One of the motivations of hidden variables models is to avoid the sudden reduction of states. It may thus be desirable to avoid the direct use of reduction in hidden variables models. In comparison, the expression (3.4) with (3.6) uses the same state before and after the measurement of $B$ and thus avoids the issue of reduction (by using the quantum mechanical product $BAB$ instead), but now
one has to explain the state preparation, namely, how to produce a desired state such as in (2.6) by an experimental procedure without referring to reduction (or its possible counter part in hidden variables models, which is not specified in Bell's construction).

We mention that essentially the same conclusion holds in connection with the conditional measurement in the explicit $d=2$ hidden variables model by Kochen and Specker~\cite{kochen}.  One may thus conclude that no satisfactory dispersion free description of the conditional measurement exists in $d=2$ noncontextual hidden variables models besides the lack of uniqueness of the expression in hidden variables space.

\section{Branching in hidden variables space}
A way to avoid the discrepancy between the expressions (3.3) and (3.6) in hidden variables space has been briefly noted elsewhere~\cite{fujikawa}. The basic idea is to unify these two expressions as
\begin{eqnarray}
&&A_{\psi_{B}}(\omega^{\prime})B_{\psi}(\omega)/\int d\omega B_{\psi}(\omega)
\end{eqnarray}
and later integrate over $\omega$ and $\omega^{\prime}$ {\em independently}; if one integrates over $\omega$ first one obtains (3.3), while  if one integrates over $\omega^{\prime}$ first one obtains (3.6).  
Namely, one may assume that each measurement opens up a new hidden variables space, which effectively incorporates the notion of reduction, and this procedure itself is somewhat analogous to the many-worlds interpretation of quantum mechanics~\cite{everett} although the real physical significance is completely different.

In this formula (5.1), the first measurement of  $B=P_{\vec{n}}$ for the state $\psi$ makes branching
\begin{eqnarray}
B_{\psi}(\omega)
\end{eqnarray}
or 
\begin{eqnarray}
\bar{B}_{\psi}(\omega)
\end{eqnarray}
with
\begin{eqnarray}
\bar{B}_{\psi}(\omega)=1-B_{\psi}(\omega).
\end{eqnarray}
In the formula (5.1), the normalization factor $1/\int d\omega B_{\psi}(\omega)$ arises to realize the quantum mechanical notion of reduction effectively in the dispersion free representation, namely, one starts anew with the measured reduced state specified by $B$: The measurement of $B$ and then the measurement of $A$ is given by ${\rm Tr}AB\rho B/{\rm Tr}B\rho$ in quantum mechanics. 

The state makes a transition in (5.2)
\begin{eqnarray}
|\psi\rangle\langle\psi|\rightarrow |\psi_{B}\rangle\langle\psi_{B}|=B
\end{eqnarray}
which can be regarded as the state preparation of $|\psi_{B}\rangle$, although in the language of {\em dispersion-free representation}. We realize the state $|\psi_{B}\rangle$ in the domain of the hidden variables
space with $B_{\psi}(\omega)=1$, $\omega \in \Lambda$. 

The subsequent measurement of the projector $A=P_{\vec{m}}$ then makes branching
\begin{eqnarray}
A_{\psi_{B}}(\omega^{\prime})B_{\psi}(\omega)
\end{eqnarray}
or 
\begin{eqnarray}
\bar{A}_{\psi_{B}}(\omega^{\prime})B_{\psi}(\omega)
\end{eqnarray}
with $\bar{A}_{\psi_{B}}(\omega^{\prime})=1-A_{\psi_{B}}(\omega^{\prime})$ if one starts with the state in (5.2).  In (5.6), the state makes a transition 
\begin{eqnarray}
|\psi_{B}\rangle\langle\psi_{B}|\rightarrow |\psi_{A}\rangle\langle\psi_{A}|=A
\end{eqnarray}
which can be regarded as the state preparation of $|\psi_{A}\rangle$, although in the language of dispersion-free representation. We have the state $|\psi_{A}\rangle$ in the domain of the hidden variables
space with $A_{\psi_{B}}(\omega^{\prime})=1$, $\omega^{\prime}\in \Lambda$. The domains with $B_{\psi}(\omega)=0$ or  $A_{\psi_{B}}(\omega^{\prime})=0$ do not contribute to (5.6). This process continues every time when one makes a new measurement of a projection operator.

The essence is that one can incorporate the notion of reduction effectively and thus the state preparation in the dispersion free representation by employing a procedure  which extends the hidden variables spaces indefinitely in the sense $\Lambda\rightarrow \Lambda\times\Lambda \rightarrow \Lambda\times\Lambda\times\Lambda \rightarrow ...\  $. 

Coming back to (5.1), 
\begin{eqnarray}
&&A_{\psi_{B}}(\omega^{\prime})B_{\psi}(\omega)/\int d\omega B_{\psi}(\omega)\nonumber\\
&&=\frac{1}{2}[1+\text{sign}(\omega^{\prime}+\frac{1}{2}|\vec{n}\cdot\vec{m}|)\text{sign}(\vec{n}\cdot\vec{m})]
\nonumber\\
&&\times\frac{1}{2}[1+\text{sign}(\omega+\frac{1}{2}|\vec{s}\cdot\vec{n}|)\text{sign}(\vec{s}\cdot\vec{n})]\frac{2}{(1+\vec{n}\cdot\vec{s})},\nonumber
\end{eqnarray}
one sees that the time sequence of the measurements is encoded, namely, the state preparation by the measurement of $B$ and then the measurement of $A$. The crucial property is that this is achieved in the {\em dispersion free} representation unlike the quantum mechanical reduction. By integrating over $\omega$ and $\omega^{\prime}$, one recovers the result of quantum mechanics
\begin{eqnarray}
&&\int d\omega^{\prime}\frac{1}{2}[1+\text{sign}(\omega^{\prime}+\frac{1}{2}|\vec{n}\cdot\vec{m}|)\text{sign}(\vec{n}\cdot\vec{m})]
\nonumber\\
&&\times\int d\omega\frac{1}{2}[1+\text{sign}(\omega+\frac{1}{2}|\vec{s}\cdot\vec{n}|)\text{sign}(\vec{s}\cdot\vec{n})]\frac{2}{(1+\vec{n}\cdot\vec{s})}\nonumber\\
&&=\frac{1}{2}[1+(\vec{n}\cdot\vec{m})].
\end{eqnarray}
In the case of multiple branching, one may divide by normalization factors such as 
\begin{eqnarray}
&&C_{\psi_{A}}(\omega^{\prime\prime})[A_{\psi_{B}}(\omega^{\prime})/\int d\omega^{\prime}A_{\psi_{B}}(\omega^{\prime})][B_{\psi}(\omega)/\int d\omega B_{\psi}(\omega)]
\end{eqnarray}
to be consistent with the notion of reduction ( and ray representation~\cite{dirac}) in quantum mechanics, if one is interested in the probability of the  final outcome $C_{\psi_{A}}(\omega^{\prime\prime})$. This insertion of the normalization factor is somewhat {\it ad hoc} but consistent with the formula (3.4). 

As for the measurement of $A$ immediately after the measurement of $A$, one has 
\begin{eqnarray}
A_{\psi_{A}}(\omega^{\prime})A_{\psi}(\omega)/\int d\omega A_{\psi}(\omega)
\end{eqnarray}
where $A_{\psi_{A}}(\omega^{\prime})=1$ has a uniform $\omega^{\prime}$
dependence. Once one projects the state to a specific state $A$, the further measurements of $A$
do not change the state any more in the present formulation of dispersion free representation.

\section{Conclusion} 

We discussed a way to avoid the non-uniqueness of the expression of conditional measurement in hidden variables space, either (3.3) or (3.6),  by employing an idea of branching in the hidden variables space. 
We discussed this problem in the 
very limited case of $d=2$ noncontextual hidden variables models. But if one extends the scheme to {\it contextual} hidden variables models, one can cover a wide variety of more realistic models and thus the present idea may become useful.  



\begin{thebibliography}{99}
\bibitem{gleason}
A. M. Gleason, J. Math. Mech. {\bf 6}, 885 (1957).
\bibitem{aspect}
A. Aspect, J. Dalibard and G. Roger, Phys. Rev. Lett. {\bf 49},
1804 (1982).
\bibitem{bell2}
J. S. Bell, Physics {\bf 1}, 195 (1965).
\bibitem{chsh}
J. F. Clauser, M. A. Horne, A. Shimony and R. A. Holt, Phys. Rev. Lett. 
{\bf 23},  888 (1969).
\bibitem{note}
It is shown that the local hidden variables model of Bell and CHSH gives $|\langle B\rangle |\leq 2\sqrt{2}$ or $|\langle B\rangle |\leq 2$ for the quantum CHSH operator $B={\bf a}\cdot {\bf \sigma}\otimes ({\bf b}+{\bf b}^{\prime})\cdot {\bf \sigma} +{\bf a}^{\prime}\cdot{\bf \sigma}\otimes ({\bf b}-{\bf b}^{\prime})\cdot{\bf \sigma}
$ depending on two different ways of evaluation, when it is applied to a  $d=4$ system of two spin-$1/2$ particles. This is due to the failure of linearity, and it shows that  the conventional CHSH inequality $|\langle B\rangle |\leq 2$ does not provide a reliable test of the $d=4$ local  non-contextual hidden variables model. The exclusion of the $d=4$ noncontextual model on the basis of Bell and CHSH inequalities is thus less definite compared to the analysis by Gleason's theorem.\\
This analysis is reported in K. Fujikawa, "Does CHSH inequality test the model of local hidden variables?" (to be published).
\bibitem{bell1}
J. S. Bell, Rev. Mod. Phys. {\bf 38}, 447 (1966).
\bibitem{kochen}
S. Kochen and E. P. Specker, J. Math. Mech. {\bf 17}, 59 (1967).
\bibitem{beltrametti}
E. G. Beltrametti and G. Gassinelli, {\em The Logic of Quantum 
Mechanics}, (Addison-Wesley Pub., 1981).
\bibitem{peres}
A. Peres, {\em Quantum Theory: Concepts and Methods }, (Kluwer Academic Pub., 1995).

\bibitem{fujikawa}
K. Fujikawa, Phys. Rev. A85 (2012) 012114; arXiv:1201.4421[quant-ph].
\bibitem{everett}
B.S. DeWitt and R.N. Graham, eds., {\em The Many-Worlds Interpretation of Quantum Mechanics}, Princeton Series in Physics, (Princeton University Press, 1973).
\bibitem{neumann}
J. von Neumann, {\em Mathematical Foundations of Quantum Mechanics},
(Princeton Univ. Press, 1955).
\bibitem{umegaki}
H. Umegaki, Tohoku Math. J. {\bf 6}, 177 (1954).
\bibitem{davies}
E. B. Davies and J. T. Lewis, Comm. Math. Phys. {\bf 17}, 239 (1970).  
\bibitem{dirac}
P.A.M. Dirac, {\em Principles of Quantum Mechanics}, (Oxford University Press, 1958).
\end{thebibliography}
\end{document}